\documentclass{article}%
\usepackage{amsmath}
\usepackage{amsfonts}
\usepackage{amssymb}
\usepackage{graphicx}%
\providecommand{\U}[1]{\protect\rule{.1in}{.1in}}

\begin{document}

\title{Entropic Dynamics\thanks{Published in Entropy \textbf{17}, 6110-6128 (2015). }}
\author{Ariel Caticha\\{\small Physics Department, University at Albany-SUNY, Albany, NY 12222, USA.}}
\date{}
\maketitle

\begin{abstract}
Entropic Dynamics is a framework in which dynamical laws are derived as an
application of entropic methods of inference. No underlying action principle
is postulated. Instead, the dynamics is driven by entropy subject to the
constraints appropriate to the problem at hand. In this paper we review three
examples of entropic dynamics. First we tackle the simpler case of a standard
diffusion process which allows us to address the central issue of the nature
of time. Then we show that imposing the additional constraint that the
dynamics be non-dissipative leads to Hamiltonian dynamics. Finally,
considerations from information geometry naturally lead to the type of
Hamiltonian that describes quantum theory.

\end{abstract}

\section{Introduction}

The laws of physics, and in particular the laws of dynamics, have
traditionally been seen as laws of nature. It is usually believed that such
laws are \emph{discovered} and that they are useful because they
\emph{reflect} reality. The reflection, imperfect though it may be, represents
a very direct relation between physics and nature. Here we explore an
alternative view in which the relation is considerably more indirect: The laws
of physics provide a framework for processing information about nature. From
this perspective physical models are mere tools that are partly discovered and
partly designed with our own very human purposes in mind. This approach is
decidedly pragmatic: when tools happen to be successful we do not say that
they are \emph{true}; we say that they are \emph{useful}.

Is there any evidence in support of such an unorthodox view? The answer is
yes. Indeed, if physics is an exercise in inference then we should expect it
to include both ontic and epistemic concepts. The ontic concepts are meant to
represent those entities in nature that are the subject of our interest. They
include the quantities, such as particle positions and field strengths, that
we want to predict, to explain, and to control. The epistemic concepts, on the
other hand, are the tools --- the probabilities, the entropies, and the
(information) geometries --- that are used to carry out our inferences. The
prominence of these epistemic elements strongly suggests that physics is not a
mere mirror of nature; instead physics is an inference framework designed by
humans for the purpose of facilitating their interactions with nature. The
founders of quantum theory --- Bohr, Heisenberg, Born, etc. --- were quite
aware of the epistemological and pragmatic elements in quantum mechanics (see
e.g., \cite{Stapp 1972}) but they wrote at a time when the tools of
quantitative epistemology --- the Bayesian and entropic methods of inference
--- had not yet been sufficiently developed.

Entropic Dynamics (ED) provides a framework for deriving dynamical laws as an
application of entropic methods. (The principle of maximum entropy as a method
for inference can be traced to E. T. Jaynes. For a pedagogical overview of
Bayesian and entropic inference and further references see \cite{Caticha
2012}.) In ED the dynamics is driven by entropy subject to the constraints
appropriate to the problem at hand. It is through these constraints that the
\textquotedblleft physics\textquotedblright\ is introduced. Such a framework
is extremely restrictive. For example, in order to adopt an epistemic view of
the quantum state $\psi$ it is not sufficient to merely assert that the
probability $|\psi|^{2}$ represents a state of knowledge; this is a good start
but it is not nearly enough. It is also necessary that the changes or updates
of the epistemic $\psi$ --- which include \emph{both} the unitary time
evolution described the Schr\"{o}dinger equation and the collapse of the wave
function during measurement --- be \emph{derived} according to the established
rules of inference. Therefore in a truly entropic dynamics we are not allowed
to postulate action principles that operate at some deeper sub-quantum level.
Instead, the goal is to derive such action principles from entropy principles
with suitably chosen constraints.

In this paper we collect and streamline results that have appeared in several
publications (see \cite{Caticha 2010a}-\cite{Caticha 2014b} and references
therein) to provide a self-contained overview of three types of ED: (1)
standard diffusion, (2) Hamiltonian dynamics, and (3) quantum mechanics. First
we tackle the case of a diffusion process which serves to address the central
concern with the nature of time. In ED \textquotedblleft
entropic\textquotedblright\ time is a relational concept introduced as a
book-keeping device designed to keep track of the accumulation of change. Time
is constructed by (a) introducing the notion of \textquotedblleft
instants\textquotedblright, (b) showing that these instants are
\textquotedblleft ordered\textquotedblright, and (c) defining a measure of the
interval that separates successive instants. The welcome new feature is that
an arrow of time is generated automatically; entropic time is intrinsically directional.

The early formulation of ED \cite{Caticha 2010a} involved assumptions about
auxiliary variables, the metric of configuration space, and the form of the
quantum potential. All these assumptions have been subsequently removed. In
\cite{Caticha 2014b} it was shown how the constraint that the dynamics be
non-dissipative leads to a generic form of Hamiltonian dynamics, with its
corresponding symplectic structure and action principle. Thus, in the context
of ED action principles are \emph{derived}; they are useful tools but they are
not fundamental.

Different Hamiltonians lead to different dynamical laws. We show how
considerations of information geometry provide the natural path to
Hamiltonians that include the correct form of \textquotedblleft quantum
potential\textquotedblright\ and lead to the Schr\"{o}dinger equation, and we
also identify the constraints that describe motion in an external
electromagnetic field.

Here we focus on the derivation of the Schr\"{o}dinger equation but the ED
approach has been applied to several other topics in quantum mechanics that
will not be reviewed here. These include the quantum measurement problem
\cite{Caticha 2014a}\cite{Johnson Caticha 2011}\cite{Johnson 2011}; momentum,
angular momentum, their uncertainty relations, and spin \cite{Nawaz Caticha
2011}\cite{Nawaz 2014}; relativistic scalar fields \cite{Ipek Caticha 2014};
the Bohmian limit \cite{Bartolomeo Caticha 2015}; and the extension to curved
spaces \cite{Nawaz et al 2015}.

There is vast literature on the attempts to reconstruct quantum mechanics and
it is inevitable that the ED approach might resemble them in one aspect or
another --- after all, in order to claim success all these approaches must
sooner or later converge to the same Schr\"{o}dinger equation. However, there
are important differences. For example, the central concern with the notion of
time makes ED significantly different from other approaches that are also
based on information theory (see \emph{e.g.}, \cite{Wootters 1981}%
-\cite{Reginatto 2013}). ED also differs from those approaches that attempt to
explain the emergence of quantum behavior as the effective statistical
mechanics of some underlying sub-quantum dynamics which might possibly include
some additional stochastic element (see \emph{e.g.}, \cite{Nelson
1985}-\cite{Grossing 2008}). Indeed, ED makes no reference to any sub-quantum
dynamics whether classical, deterministic, or stochastic.

\section{Entropic Dynamics}

As with other applications of entropic methods, to derive dynamical laws we
must first specify the microstates that are the subject of our inference ---
the subject matter --- and then we must specify the prior probabilities and
the constraints that represent the information that is relevant to our
problem. (See \emph{e.g.}, \cite{Caticha 2012}.) We consider $N$ particles
living in a flat Euclidean space $\mathbf{X}$ with metric $\delta_{ab}$. We
assume that the particles have definite positions $x_{n}^{a}$ and it is their
unknown values that we wish to infer.\footnote{In this work ED is developed as
a model for the quantum mechanics of particles. The same framework can be
deployed to construct models for the quantum mechanics of fields, in which
case it is the fields that are objectively \textquotedblleft
real\textquotedblright\ and have well-defined albeit unknown values.\cite{Ipek
Caticha 2014}} (The index $n$ $=1\ldots N$ denotes the particle and $a=1,2,3$
the spatial coordinates.) For $N$ particles the configuration space is
$\mathbf{X}_{N}=\mathbf{X}\times\ldots\times\mathbf{X}$.

The assumption that the particles have definite positions is in flat
contradiction with the standard Copenhagen notion that quantum particles
acquire definite positions only as a result of a measurement. For example, in
the ED description of the double slit experiment we do not know which slit the
quantum particle goes through but it most definitely goes through either one
or the other.

We do not explain why motion happens but, given the information that it does,
our task is to produce an estimate of what short steps we can reasonably
expect. The next assumption is dynamical: we assume that the particles follow
trajectories that are continuous. This means that displacements over finite
distances can be analyzed as the accumulation of many infinitesimally short
steps and our first task is to find the transition probability density
$P(x^{\prime}|x)$ for a single short step from a given initial $x\in
\mathbf{X}_{N}$ to an unknown neighboring $x^{\prime}\in\mathbf{X}_{N}$. Later
we will determine how such short steps accumulate to yield a finite displacement.

To find $P(x^{\prime}|x)$ we maximize the (relative) entropy,
\begin{equation}
\mathcal{S}[P,Q]=-\int dx^{\prime}\,P(x^{\prime}|x)\log\frac{P(x^{\prime}%
|x)}{Q(x^{\prime}|x)}~. \label{Sppi}%
\end{equation}
To simplify the notation in all configuration space integrals we write
$d^{3N}x=dx$, $d^{3N}x^{\prime}=dx^{\prime}$, and so on. $Q(x^{\prime}|x)$ is
the prior probability. It expresses our beliefs --- or more precisely, the
beliefs of an ideally rational agent --- \emph{before} any information about
the motion is taken into account. The physically relevant information about
the step is expressed in the form of constraints imposed on $P(x^{\prime}|x)$
--- this is the stage at which the physics is introduced.

\paragraph*{The prior --}

We adopt a prior $Q(x^{\prime}|x)$ that represents a state of extreme
ignorance: knowledge of the initial position $x$ tells us nothing about
$x^{\prime}$. Such ignorance is expressed by assuming that $Q(x^{\prime
}|x)dx^{\prime}$ is proportional to the volume element in $\mathbf{X}_{N}$.
Since the space $\mathbf{X}_{N}$ is flat and a mere proportionality constant
has no effect on the entropy maximization we can set $Q(x^{\prime}%
|x)=1$.\footnote{Uniform non-normalizable priors are mathematically
problematic. For us this is only a mild annoyance that is easily evaded by
adopting a physically reasonable normalizable prior. The idea is that for a
microscopic particle such as an electron a \textquotedblleft
uniform\textquotedblright\ prior is any distribution that is essentially flat
over macroscopic scales. For example we can use a Gaussian centered at $x$
with a macroscopically large standard deviation.}

\paragraph*{The constraints --}

The first piece of information is that motion is continuous --- motion
consists of a succession of infinitesimally short steps. Each individual
particle $n$ will take a short step from $x_{n}^{a}$ to $x_{n}^{\prime
a}=x_{n}^{a}+\Delta x_{n}^{a}$ and we require that for each particle the
expected squared displacement,
\begin{equation}
\langle\Delta x_{n}^{a}\Delta x_{n}^{b}\rangle\delta_{ab}=\kappa_{n}%
~,\qquad(n=1\ldots N)~ \label{kappa n}%
\end{equation}
takes some small value $\kappa_{n}$. Infinitesimally short steps are obtained
by taking the limit $\kappa_{n}\rightarrow0$. We will assume each $\kappa_{n}$
to be independent of $x$ to reflect the translational symmetry of $\mathbf{X}%
$. In order to describe non-identical particles we assume that the value of
$\kappa_{n}$ depends on the particle index $n$.

The $N$ constraints (\ref{kappa n}) treat the particles as statistically
independent and their accumulation eventually leads to a completely isotropic
diffusion. But we know that particles can become correlated and even become
entangled. We also know that motion is not normally isotropic; once particles
are set in motion they tend to persist in it. This information is introduced
through one additional constraint involving a \textquotedblleft
drift\textquotedblright\ potential $\phi(x)$ that is a function in
configuration space, $x\in\mathbf{X}_{N}$. We impose that the expected
displacements $\left\langle \Delta x_{n}^{a}\right\rangle $ along the
direction of the gradient of $\phi$ satisfy
\begin{equation}
\langle\Delta x^{A}\rangle\partial_{A}\phi=\sum\limits_{n=1}^{N}\left\langle
\Delta x_{n}^{a}\right\rangle \frac{\partial\phi}{\partial x_{n}^{a}}%
=\kappa^{\prime}~, \label{kappa prime}%
\end{equation}
where capitalized indices such as $A=(n,a)$ include both the particle index
and its spatial coordinate; $\partial_{A}=\partial/\partial x^{A}%
=\partial/\partial x_{n}^{a}$; and $\kappa^{\prime}$ is another small but for
now unspecified position-independent constant.

The introduction of the drift potential $\phi$ will not be justified at this
point. The idea is that we can make progress by identifying the constraints
even when their physical origin remains unexplained. This situation is not
unlike classical mechanics where identifying the forces is useful even in
situations where their microscopic origin is not understood. We do however
make two brief comments. First, in section 9 we shall see that ED in an
external electromagnetic field is described by constraints that are formally
similar to (\ref{kappa prime}). There we shall show that the effects of the
drift potential $\phi$ and the electromagnetic vector potential $A_{a}$ are
intimately related --- a manifestation of gauge symmetry --- suggesting that
whatever $\phi$ might be, it is as \textquotedblleft real\textquotedblright%
\ as $A_{a}$. The second comment is that elsewhere, in the context of a
particle with spin, we will see that the drift potential can be given a
natural geometric interpretation as an angular variable. This imposes the
additional condition that the integral of $\phi$ over any closed loop is
quantized, $%
{\textstyle\oint}
d\phi=2\pi\nu$ where $\nu$ is an integer.

Maximizing $\mathcal{S}[P,Q]$ in (\ref{Sppi}) subject to the $N+2$ constraints
(\ref{kappa n}), (\ref{kappa prime}) plus normalization yields a Gaussian
distribution,
\begin{equation}
P(x^{\prime}|x)=\frac{1}{\zeta}\exp%
{\displaystyle\sum\nolimits_{n}}
\left[  -\frac{1}{2}\alpha_{n}\,\Delta x_{n}^{a}\Delta x_{n}^{b}\delta
_{ab}+\alpha^{\prime}\Delta x_{n}^{a}\frac{\partial\phi}{\partial x_{n}^{a}%
}\right]  ~, \label{Prob xp/x a}%
\end{equation}
where $\zeta=\zeta(x,\alpha_{n},\alpha^{\prime})$ is a normalization constant
and the Lagrange multipliers $\alpha_{n}$ and $\alpha^{\prime}$ are determined
from
\begin{equation}
\partial\log\zeta/\partial\alpha_{n}=-\kappa_{n}/2\quad\text{and\quad}%
\partial\log\zeta/\partial\alpha^{\prime}=\kappa^{\prime}~.
\end{equation}
The distribution $P(x^{\prime}|x)$ is conveniently rewritten as
\begin{equation}
P(x^{\prime}|x)=\frac{1}{Z}\exp\left[  -\frac{1}{2}%
{\displaystyle\sum\nolimits_{n}}
\alpha_{n}\,\delta_{ab}(\Delta x_{n}^{a}-\langle\Delta x_{n}^{a}%
\rangle)(\Delta x_{n}^{b}-\langle\Delta x_{n}^{b}\rangle)\right]  ~,
\label{Prob xp/x b}%
\end{equation}
where $Z$ is a new normalization constant. A generic displacement $\Delta
x_{n}^{a}=x_{n}^{\prime a}-x_{n}^{a}$ can be expressed as an expected drift
plus a fluctuation,
\begin{equation}
\Delta x_{n}^{a}=\left\langle \Delta x_{n}^{a}\right\rangle +\Delta w_{n}%
^{a}\,\,,\quad\text{where\quad}\langle\Delta x_{n}^{a}\rangle=\frac
{\alpha^{\prime}}{\alpha_{n}}\delta^{ab}\frac{\partial\phi}{\partial x_{n}%
^{b}}~, \label{ED drift}%
\end{equation}%
\begin{equation}
\left\langle \Delta w_{n}^{a}\right\rangle =0\quad\text{and}\quad\langle\Delta
w_{n}^{a}\Delta w_{n}^{b}\rangle=\frac{1}{\alpha_{n}}\delta^{ab}~.
\label{ED fluctuations}%
\end{equation}
From these equations we can get a first glimpse into the meaning of the
multipliers $\alpha_{n}$ and $\alpha^{\prime}$. For very short steps, as
$\alpha_{n}\rightarrow\infty$, the fluctuations become dominant: the drift is
$\langle\Delta x_{n}^{a}\rangle\sim O(\alpha_{n}^{-1})$ while $\Delta
w_{n}^{a}\sim O(\alpha_{n}^{-1/2})$. This implies that, as in Brownian motion,
the trajectory is continuous but not differentiable. In the ED approach a
particle has a definite position but its velocity, the tangent to the
trajectory, is completely undefined. We can also see that the effect of
$\alpha^{\prime}$ is to enhance or suppress the magnitude of the drift
relative to the fluctuations --- a subject that is discussed in detail in
\cite{Bartolomeo Caticha 2015}. However, for our current purposes we can
absorb $\alpha^{\prime}$ into the so far unspecified drift potential,
$\alpha^{\prime}\phi\rightarrow\phi$, which amounts to setting $\alpha
^{\prime}=1$.

\section{Entropic time}

The goal is to derive dynamical laws as an application of inference methods
but the latter make no reference to time. Therefore, additional assumptions
are needed to specify what constitutes \textquotedblleft
time\textquotedblright\ in our framework.

\emph{The foundation to any notion of time is dynamics.} We must first
identify a notion of \textquotedblleft instant\textquotedblright\ that
properly takes into account the inferential nature of entropic dynamics. Time
is then is constructed as a device to keep track of change or, more
explicitly, of the accumulation of repeated small changes.

\subsection{Time as an ordered sequence of instants}

Entropic dynamics is generated by the short-step transition probability
$P(x^{\prime}|x)$, eq.(\ref{Prob xp/x b}). In the $i$th step, which takes the
system from $x=x_{i-1}$ to $x^{\prime}=x_{i}$, both $x_{i-1}$ and $x_{i}$ are
unknown. Integrating the joint probability, $P(x_{i},x_{i-1})$, over $x_{i-1}$
gives
\begin{equation}
P(x_{i})=\int dx_{i-1}P(x_{i},x_{i-1})=\int dx_{i-1}P(x_{i}|x_{i-1}%
)P(x_{i-1})~. \label{CK a}%
\end{equation}
These equations involve no assumptions; they are true by virtue of the laws of
probability. However, if $P(x_{i-1})$ happens to be the probability of
different values of $x_{i-1}$ \emph{at an \textquotedblleft
instant\textquotedblright\ labelled }$t$, then we can interpret $P(x_{i})$ as
the probability of values of $x_{i}$ \emph{at the next \textquotedblleft
instant\textquotedblright} which we will label $t^{\prime}$. Accordingly, we
write $P(x_{i-1})=\rho(x,t)$ and $P(x_{i})=\rho(x^{\prime},t^{\prime})$ so
that
\begin{equation}
\rho(x^{\prime},t^{\prime})=\int dx\,P(x^{\prime}|x)\rho(x,t)~, \label{CK b}%
\end{equation}
Nothing in the laws of probability leading to eq.(\ref{CK a}) forces the
interpretation (\ref{CK b}) on us --- this is an independent assumption about
what constitutes time in our model. We use eq.(\ref{CK b}) to define what we
mean by an instant:\emph{\ if the distribution }$\rho(x,t)$\emph{\ refers to
one instant }$t$\emph{, then the distribution }$\rho(x^{\prime},t^{\prime}%
)$\emph{\ generated by }$P(x^{\prime}|x)$\emph{ defines what we mean by the
\textquotedblleft next\textquotedblright\ instant }$t^{\prime}$. The iteration
of this process defines the dynamics: entropic time is constructed instant by
instant: $\rho(t^{\prime})$ is constructed from $\rho(t)$, $\rho
(t^{\prime\prime})$ is constructed from $\rho(t^{\prime})$, and so on.

The inferential nature of the construction can be phrased more explicitly.
Once we have decided on the relevant information necessary for predicting
future behavior --- the distributions $\rho(x,t)$ and $P(x^{\prime}|x)$ --- we
can say that this information defines what we mean by an \textquotedblleft
instant\textquotedblright. Furthermore, eq.(\ref{CK b}) shows that

\begin{description}
\item[\qquad] \emph{Time is }designed\emph{ in such a way that\ given the
present the future is independent of the past. }
\end{description}

\noindent An equation such as (\ref{CK b}) is commonly employed to define
Markovian behavior in which case it is sometimes known as the
Chapman-Kolmogorov equation. Markovian processes are such that once an
external notion of time is given, defined perhaps by an external clock, the
specification of the state of the system at time $t$ is sufficient to fully
determine its state after time $t$ --- no additional information about times
prior to $t$ is needed. It should be emphasized that we do not make a
Markovian assumption. We are concerned with a different problem: we do not use
(\ref{CK b}) to define a Markovian process \emph{in an already existing
background time}; we use it to construct time itself.

\subsection{The arrow of entropic time}

The notion of time constructed according to eq.(\ref{CK b}) incorporates an
intrinsic directionality: there is an absolute sense in which $\rho(x,t)$\ is
prior and $\rho(x^{\prime},t^{\prime})$\ is posterior. To construct the
time-reversed evolution we just write
\begin{equation}
\rho(x,t)=%
{\textstyle\int}
dx^{\prime}\,P(x|x^{\prime})\rho(x^{\prime},t^{\prime})\,,
\end{equation}
where according to the rules of probability theory $P(x|x^{\prime})$ is
related to $P(x^{\prime}|x)$ in eq.(\ref{Prob xp/x b}) by Bayes' theorem,
\begin{equation}
P(x|x^{\prime})=\frac{\rho(x,t)}{\rho(x^{\prime},t^{\prime})}P(x^{\prime}|x)~.
\label{Bayes thm}%
\end{equation}
Note that this is not a mere exchange of primed and unprimed quantities:
$P(x^{\prime}|x)$ is the Gaussian distribution, eq.(\ref{Prob xp/x b}),
obtained using the maximum entropy method; in contrast, the time-reversed
$P(x|x^{\prime})$ obtained using Bayes' theorem, eq.(\ref{Bayes thm}), will
not in general be Gaussian. The asymmetry between the inferential past and the
inferential future is traced to the asymmetry between priors and posteriors.

The subject of the arrow of time has a vast literature (see \emph{e.g.}
\cite{Price 1996}\cite{Zeh 2002}). The puzzle has been how to explain the
asymmetric arrow from underlying symmetric laws of nature. The solution
offered by ED is that there are no symmetric underlying laws and the asymmetry
is the inevitable consequence of entropic inference. From the point of view of
ED the challenge is not to explain the arrow of time, but rather the reverse,
how to explain the emergence of symmetric laws within an intrinsically
asymmetric entropic framework. As we shall see below the derived laws of
physics --- e.g., the Schr\"{o}dinger equation --- are time-reversible but
\emph{entropic time itself only f{}lows forward}.

\subsection{Duration: a convenient time scale}

To complete the model of entropic time we need to specify the interval $\Delta
t$ between successive instants. The basic criterion is that \emph{duration is
defined so that motion looks simple.} From eqs.(\ref{ED drift}) and
(\ref{ED fluctuations}) for short steps (large $\alpha_{n}$) the motion is
dominated by the f{}luctuations $\Delta w_{n}$. Therefore specifying $\Delta
t$ amounts to specifying the multipliers $\alpha_{n}$ in terms of $\Delta t$.

To start we appeal to symmetry: in order that the f{}luctuations $\left\langle
\Delta w_{n}^{a}\Delta w_{n}^{b}\right\rangle $ ref{}lect the symmetry of
translations in space and time --- a time that just like Newton's flows
\textquotedblleft equably\ everywhere and everywhen\textquotedblright\ --- we
choose $\alpha_{n}$ to be independent of $x$ and $t$, and we choose $\Delta t$
so that $\alpha_{n}\propto1/\Delta t$. More explicitly, we write
\begin{equation}
\alpha_{n}=\frac{m_{n}}{\eta}\frac{1}{\Delta t}~, \label{alpha n}%
\end{equation}
where the proportionality constants have been written in terms of some
particle-specific constants $m_{n}$, which will eventually be identified with
the particle masses, and an overall constant $\eta$ that fixes the units of
the $m_{n}$s relative to the units of time and will later be regraduated into
$\hbar$.

Before discussing the implications of the choice (\ref{alpha n}) it is useful
to consider the geometry of the $N$-particle configuration space,
$\mathbf{X}_{N}$.

\section{The information metric of configuration space}

We have assumed that the geometry of the single particle spaces $\mathbf{X}$
is described by the Euclidean metric $\delta_{ab}$. We can expect that the
$N$-particle configuration space, $\mathbf{X}_{N}=\mathbf{X}\times\ldots
\times\mathbf{X}$, will also have a flat geometry, but the relative
contribution of different particles to the metric remains undetermined. Should
very massive particles contribute the same as very light particles? The answer
is provided by information geometry.

To each point $x\in\mathbf{X}_{N}$ there corresponds a probability
distribution $P(x^{\prime}|x)$. Therefore $\mathbf{X}_{N}$ is a statistical
manifold and up to an arbitrary global scale factor its geometry is uniquely
determined by the information metric,%
\begin{equation}
\gamma_{AB}=C\int dx^{\prime}\,P(x^{\prime}|x)\frac{\partial\log P(x^{\prime
}|x)}{\partial x^{A}}\frac{\partial\log P(x^{\prime}|x)}{\partial x^{B}}~,
\label{gamma C}%
\end{equation}
where $C$ is an arbitrary positive constant (see \emph{e.g.}, \cite{Caticha
2012}). A straightforward substitution of eqs.(\ref{Prob xp/x b}) and
(\ref{alpha n}) into eq.(\ref{gamma C}) in the limit of short steps
($\alpha_{n}\rightarrow\infty$) yields
\begin{equation}
\gamma_{AB}=\frac{Cm_{n}}{\eta\Delta t}\delta_{nn^{\prime}}\,\delta_{ab}%
=\frac{Cm_{n}}{\eta\Delta t}\delta_{AB}~. \label{gamma AB}%
\end{equation}
We see that $\gamma_{AB}$ diverges as $\Delta t\rightarrow0$. The reason for
this is not hard to find. As the Gaussian distributions $P(x^{\prime}|x)$ and
$P(x^{\prime}|x+\Delta x)$ become more sharply peaked and it is easier to
distinguish one from the other which translates into a greater information
distance, $\gamma_{AB}\rightarrow\infty$. In order to define a distance that
remains meaningful for arbitrarily small $\Delta t$ it is convenient to choose
$C\propto\Delta t$. In what follows the metric tensor will always appear in
combinations such as $\gamma_{AB}\Delta t/C$. It is therefore convenient to
define the \textquotedblleft mass\textquotedblright\ tensor,
\begin{equation}
m_{AB}=\frac{\eta\Delta t}{C}\gamma_{AB}=m_{n}\delta_{AB}~. \label{mass a}%
\end{equation}
Its inverse,
\begin{equation}
m^{AB}=\frac{C}{\eta\Delta t}\gamma^{AB}=\frac{1}{m_{n}}\delta^{AB}~,
\label{mass b}%
\end{equation}
is called the \textquotedblleft diffusion\textquotedblright\ tensor.

We can now summarize our results so far. The choice (\ref{alpha n}) of the
multipliers $\alpha_{n}$ simplifies the dynamics: $P(x^{\prime}|x)$
in\ (\ref{Prob xp/x b}) is a standard Wiener process. A generic displacement,
eq.(\ref{ED drift}), is
\begin{equation}
\Delta x^{A}=b^{A}\Delta t+\Delta w^{A}~, \label{Delta x}%
\end{equation}
where $b^{A}(x)$ is the drift velocity,
\begin{equation}
\langle\Delta x^{A}\rangle=b^{A}\Delta t\quad\text{with}\quad b^{A}=\frac
{\eta}{m_{n}}\delta^{AB}\partial_{B}\phi=\eta m^{AB}\partial_{B}\phi~,
\label{drift velocity}%
\end{equation}
and the fluctuations $\Delta w^{A}$ are such that
\begin{equation}
\langle\Delta w^{A}\rangle=0\quad\text{and}\quad\langle\Delta w^{A}\Delta
w^{B}\rangle=\frac{\eta}{m_{n}}\delta^{AB}\Delta t=\eta m^{AB}\Delta t~.
\label{fluc}%
\end{equation}

We are now ready to comment on the implications of the choice of time scale
$\Delta t$ and of multipliers $\alpha_{n}$, eq.(\ref{alpha n}). The first
remark is on the nature of clocks: In Newtonian mechanics the prototype of a
clock is the free particle and time is defined so as to simplify the motion of
free particles --- they move equal distances in equal times. In ED the
prototype of a clock is a free particle too --- for sufficiently short times
all particles are free --- and time here is also defined to simplify the
description of their motion: the particle undergoes equal fluctuations in
equal times.

The second remark is on the nature of mass. The particle-specific constants
$m_{n}$ will, in due course, be called `mass' and eq.(\ref{fluc}) provides
their interpretation: mass is an inverse measure of fluctuations. Thus, up to
overall constants the metric of configuration space is the mass tensor and its
inverse is the diffusion tensor. In standard QM there are two mysteries:
\textquotedblleft Why quantum fluctuations?\textquotedblright\ and
\textquotedblleft What is mass?\textquotedblright. ED offers some progress in
that instead of two mysteries there is only one. Fluctuations and mass are two
sides of the same coin.

Finally we note the formal similarity to Nelson's stochastic mechanics
\cite{Nelson 1985}. The similarity is to be expected --- all theories that
converge on the Schr\"{o}dinger equation must at some point become formally
similar --- but our epistemic interpretation differs radically from Nelson's
ontic interpretation and avoids the difficulties discussed in \cite{Nelson
2012}.

\section{Diffusive dynamics}

Equation (\ref{CK b}) is the dynamical equation for the evolution of
$\rho(x,t)$. It is written in integral form but it can be written in
differential form as a Fokker-Planck (FP) equation (see \emph{e.g.},
\cite{Caticha 2012})
\begin{equation}
\partial_{t}\rho=-\partial_{A}\left(  b^{A}\rho\right)  +\frac{1}{2}\eta
m^{AB}\partial_{A}\partial_{B}\rho~, \label{FP a}%
\end{equation}
or equivalently as a continuity equation,
\begin{equation}
\partial_{t}\rho=-\partial_{A}\left(  \rho v^{A}\right)  ~, \label{FP b}%
\end{equation}
where $v^{A}$ is the velocity of the probability flow or \emph{current
velocity},
\begin{equation}
v^{A}=b^{A}+u^{A}\quad\text{and}\quad u^{A}=-\eta m^{AB}\partial_{B}\log
\rho^{1/2}~
\end{equation}
is the \emph{osmotic velocity }--- it represents the tendency for probability
to flow down the density gradient. Since both $b^{A}$ and $u^{A}$ are
gradients, it follows that the current velocity is a gradient too,%
\begin{equation}
v^{A}=m^{AB}\partial_{B}\Phi\quad\text{where}\quad\Phi=\eta(\phi-\log
\rho^{1/2})~. \label{curr}%
\end{equation}
The FP equation
\begin{equation}
\partial_{t}\rho=-\partial_{A}\left(  \rho m^{AB}\partial_{B}\Phi\right)  ~,
\label{FP c}%
\end{equation}
can be conveniently rewritten in the alternative form
\begin{equation}
\partial_{t}\rho=\frac{\delta\tilde{H}}{\delta\Phi}~, \label{Hamilton a}%
\end{equation}
for some suitably chosen functional $\tilde{H}[\rho,\Phi]$. It is easy to
check that the appropriate functional $\tilde{H}$ is%
\begin{equation}
\tilde{H}[\rho,\Phi]=\int dx\,\frac{1}{2}\rho m^{AB}\partial_{A}\Phi
\partial_{B}\Phi+F[\rho]~, \label{Hamiltonian a}%
\end{equation}
where the integration constant $F[\rho]$ is some unspecified functional of
$\rho$.

With these results we have demonstrated that a specific form of dynamics --- a
standard diffusion process --- can be derived from principles of entropic
inference. This diffusive dynamics can be written in different but equivalent
ways --- equations eq.(\ref{FP a}), (\ref{FP b}), (\ref{FP c}), and
(\ref{Hamilton a}) are all equivalent. Next we turn our attention to other
forms of dynamics such as quantum or classical mechanics which require a
somewhat different choice of constraints.

\section{Hamiltonian dynamics}

The previous discussion has led us to a \emph{standard} diffusion in which the
density $\rho$ evolves under the influence of some externally fixed drift
potential $\phi$. However, in quantum dynamics we require a second degree of
freedom, the phase of the wave function. The extra degree of freedom is
introduced into ED by replacing the constraint of a fixed drift potential
$\phi$ by an evolving constraint in which at each time step the potential
$\phi$ is readjusted in response to the evolving $\rho$.

To find the appropriate readjustment of $\phi$ we borrow an idea of Nelson's
\cite{Nelson 1979} and impose that the potential $\phi$ be updated in such a
way that a certain functional, later called \textquotedblleft
energy\textquotedblright, remains constant. The next challenge is to identify
the appropriate functional form of this energy but before this we make two remarks.

The standard procedure in mechanics is to derive the conservation of energy
from the invariance of the action under time translations but here we do not
have an action yet. The logic of our derivation runs in the opposite
direction: we first identify the conservation of an energy as the piece of
information that is relevant to our inferences and from it we derive
Hamilton's equations and their associated action principle.

Imposing energy\ conservation appears to be natural because it agrees with our
classical preconceptions of what mechanics is like. But ED is not at all like
classical mechanics. Indeed, eq.(\ref{Delta x}) is the kind of equation (a
Langevin equation) that characterizes a Brownian motion in the limit of
\emph{infinite} friction. Therefore in the ED approach to quantum theory
particles seem to be subject to infinite friction while suffering zero
dissipation. Such a strange dynamics can hardly be called `mechanics' much
less `classical'.

\paragraph{The ensemble Hamiltonian --}

The energy functional that codifies the correct constraint is of the form
(\ref{Hamiltonian a}). We therefore impose that, irrespective of the initial
conditions, the potential $\phi$ will be updated in such a way that the
functional $\tilde{H}[\rho,\Phi]$ in (\ref{Hamiltonian a}) is always
conserved,
\begin{equation}
\frac{d\tilde{H}}{dt}=\int dx\,\left[  \frac{\delta\tilde{H}}{\delta\Phi
}\partial_{t}\Phi+\frac{\delta\tilde{H}}{\delta\rho}\partial_{t}\rho\right]
=0~.
\end{equation}
Using eq.(\ref{Hamilton a}) we get
\begin{equation}
\frac{d\tilde{H}}{dt}=\int dx\,\left[  \partial_{t}\Phi+\frac{\delta\tilde{H}%
}{\delta\rho}\right]  \partial_{t}\rho=0~. \label{dHdt}%
\end{equation}
We require that $\tilde{H}={}$const.$~$for arbitrary choices of the initial
values of $\rho$ and $\Phi$. From eq.(\ref{FP c}) we see that this amounts to
imposing $d\tilde{H}/dt=0$ for arbitrary choices of $\partial_{t}\rho$.
Therefore the requirement that $\tilde{H}$ be conserved for arbitrary initial
conditions amounts to imposing that
\begin{equation}
\partial_{t}\Phi=-\frac{\delta\tilde{H}}{\delta\rho}~. \label{Hamilton b}%
\end{equation}
Equations (\ref{Hamilton a}) and (\ref{Hamilton b}) have the form of a
canonically conjugate pair of Hamilton's equations. The conserved functional
$\tilde{H}[\rho,\Phi]$ in (\ref{Hamiltonian a}) will be called the ensemble
Hamiltonian. We conclude that non-dissipative ED leads to Hamiltonian dynamics.

\paragraph*{The action, Poisson brackets, etc. --}

The field $\rho$ is a generalized coordinate and $\Phi$ is its canonical
momentum. Eq.(\ref{Hamilton b}) leads to a generalized Hamilton-Jacobi
equation,%
\begin{equation}
\partial_{t}\Phi=-\frac{1}{2}m^{AB}\partial_{A}\Phi\partial_{B}\Phi
-\frac{\delta F}{\delta\rho}~. \label{HJ}%
\end{equation}
Now that we have Hamilton's equations, (\ref{Hamilton a}) and
(\ref{Hamilton b}), we can invert the usual procedure and \emph{construct} an
action principle from which they can be derived. Define the differential
\begin{equation}
\delta A=\int dt\int dx\left[  \left(  \partial_{t}\rho-\,\frac{\delta
\tilde{H}}{\delta\Phi}\right)  \delta\Phi-\left(  \partial_{t}\Phi
+\frac{\delta\tilde{H}}{\delta\rho}\right)  \delta\rho\right]
\end{equation}
and then integrate to get the action
\begin{equation}
A[\rho,\Phi]=\int dt\left(  \int dx\,\Phi\dot{\rho}-\tilde{H}[\rho
,\Phi]\right)  ~.
\end{equation}
By construction, imposing $\delta A=0$ leads to (\ref{Hamilton a}) and
(\ref{Hamilton b}).

The time evolution of any arbitrary functional $f[\rho,\Phi]$ is given by a
Poisson bracket,
\begin{equation}
\frac{d}{dt}f[\rho,\Phi]=\int dx\left[  \frac{\delta f}{\delta\rho}%
\frac{\delta\tilde{H}}{\delta\Phi}-\frac{\delta f}{\delta\Phi}\frac
{\delta\tilde{H}}{\delta\rho}\right]  =\{f,\tilde{H}\}~,
\end{equation}
which shows that the ensemble Hamiltonian $\tilde{H}$ is the generator of time
evolution. Similarly, under a spatial displacement $\varepsilon^{a}$ the
change in $f[\rho,\Phi]$ is
\begin{equation}
\delta_{\varepsilon}f[\rho,\Phi]=\left\{  f,\tilde{P}_{a}\varepsilon
^{a}\right\}
\end{equation}
where
\begin{equation}
\tilde{P}_{a}=\int d^{3N}x\,\rho\sum_{n}\frac{\partial\Phi}{\partial x_{n}%
^{a}}=\int d^{3N}x\,\rho\frac{\partial\Phi}{\partial X^{a}}%
\end{equation}
is interpreted as the expectation of the total momentum, and $X^{a}$ are the
coordinates of the center of mass,
\begin{equation}
X^{a}=\frac{1}{M}\sum_{n}m_{n}x_{n}^{a}~.
\end{equation}

\paragraph*{A Schr\"{o}dinger-like equation --}

We can always combine $\rho$ and $\Phi$ to define the family of complex
functions,
\begin{equation}
\Psi_{k}=\rho^{1/2}\exp(ik\Phi/\eta)\,,~ \label{psi k}%
\end{equation}
where $k$ is some arbitrary positive constant. Then the two coupled equations
(\ref{Hamilton a}) and (\ref{Hamilton b}) can be written as a single complex
Schr\"{o}dinger-like\ equation,
\begin{equation}
i\frac{\eta}{k}\partial_{t}\Psi_{k}=-\frac{1}{2}\frac{\eta^{2}}{k^{2}}%
\,m^{AB}\partial_{A}\partial_{B}\Psi_{k}+\frac{1}{2}\frac{\eta^{2}}{k^{2}%
}\,m^{AB}\frac{\partial_{A}\partial_{B}|\Psi_{k}|}{|\Psi_{k}|}\Psi_{k}%
+\frac{\delta F}{\delta\rho}\Psi_{k}~. \label{sch a}%
\end{equation}
The reason for the parameter $k$ will become clear shortly, but even at this
stage we can already anticipate that $\eta/k$ will play the role of $\hbar$.

\section{Information geometry and the Quantum Potential}

Different choices of the functional $F[\rho]$ in (\ref{Hamiltonian a}) lead to
different dynamics. Earlier we invoked information geometry, eq.(\ref{gamma C}%
), to define the metric $m_{AB}$ induced in configuration space by the
transition probabilities $P(x^{\prime}|x)$. To motivate the particular choice
of the functional $F[\rho]$ that leads to quantum theory we appeal to
information geometry once again.

Consider the family of distributions $\rho(x|\theta)$ that are generated from
a distribution $\rho(x)$ by pure translations by a vector $\theta^{A}$,
$\rho(x|\theta)=\rho(x-\theta)$. The extent to which $\rho(x|\theta)$ can be
distinguished from the slightly displaced $\rho(x|\theta+d\theta)$ or,
equivalently, the information distance between $\theta^{A}$ and $\theta
^{A}+d\theta^{A}$, is given by
\begin{equation}
d\ell^{2}=g_{AB}d\theta^{A}d\theta^{B} \label{dl^2}%
\end{equation}
where%
\begin{equation}
g_{AB}(\theta)=\int dx\frac{1}{\rho(x-\theta)}\frac{\partial\rho(x-\theta
)}{\partial\theta^{A}}\frac{\partial\rho(x-\theta)}{\partial\theta^{B}}~.
\end{equation}
Changing variables $x-\theta\rightarrow x$ yields%
\begin{equation}
g_{AB}(\theta)=\int dx\frac{1}{\rho(x)}\frac{\partial\rho(x)}{\partial x^{A}%
}\frac{\partial\rho(x)}{\partial x^{B}}=I_{AB}[\rho]~. \label{Fisher}%
\end{equation}

\paragraph*{The functional $F[\rho]$ --}

The simplest choice of functional $F[\rho]$ is linear in $\rho$, $F[\rho]=\int
dx\,\rho V$, where $V(x)$ is some function that will be recognized as the
familiar scalar potential. Since ED aims to derive the laws of physics from a
framework for inference it is natural to expect that the Hamiltonian might
also contain terms that are of a purely informational nature. We have
identified two such tensors: one is the information metric of configuration
space $\gamma_{AB}\propto m_{AB}$, another is $I_{AB}[\rho]$. The simplest
nontrivial scalar that can be constructed from them is the trace $m^{AB}%
I_{AB}$. This suggests%
\begin{equation}
F[\rho]=\xi m^{AB}I_{AB}[\rho]+\int dx\,\rho V~,~ \label{QP a}%
\end{equation}
where $\xi>0$ is a constant that controls the relative strength of the two
contributions. The term $m^{AB}I_{AB}$ is sometimes called the
\textquotedblleft quantum\textquotedblright\ or the \textquotedblleft
osmotic\textquotedblright\ potential.\footnote{The relation between the
quantum potential and the Fisher information was pointed out in
\cite{Reginatto 1998}.} From eq.(\ref{Fisher}) we see that $m^{AB}I_{AB}$ is a
contribution to the energy such that those states that are more smoothly
spread out tend to have lower energy. The case $\xi<0$ leads to instabilities
and is therefore excluded; the case $\xi=0$ leads to a qualitatively different
theory and will be discussed elsewhere. \cite{Bartolomeo Caticha 2015}

With this choice of $F[\rho]$ the generalized HJ equation becomes
\begin{equation}
-\partial_{t}\Phi=\frac{1}{2}m^{AB}\partial_{A}\Phi\partial_{B}\Phi+V-4\xi
m^{AB}\frac{\partial_{A}\partial_{B}\rho^{1/2}}{\rho^{1/2}}~. \label{HJb}%
\end{equation}

\section{The Schr\"{o}dinger equation}

Substituting eq.(\ref{QP a}) into (\ref{sch a}) gives a Schr\"{o}dinger-like
equation,%
\begin{equation}
i\frac{\eta}{k}\partial_{t}\Psi_{k}=-\frac{\eta^{2}}{2k^{2}}m^{AB}\partial
_{A}\partial_{B}\Psi_{k}+V\Psi_{k}+\left(  \frac{\eta^{2}}{2k^{2}}%
-4\xi\right)  m^{AB}\frac{\partial_{A}\partial_{B}|\Psi_{k}|}{|\Psi_{k}|}%
\Psi_{k}~, \label{sch b}%
\end{equation}
the beauty of which is severely marred by the non-linear last term.

\paragraph*{Regraduation --}

We can now make good use of the freedom afforded by the arbitrary constant
$k$. Since the physics is fully described by $\rho$ and $\Phi$ the different
choices of $k$ in $\Psi_{k}$ all describe the same theory. Among all these
equivalent descriptions it is clearly to our advantage to pick the $k$ that is
most \emph{convenient} --- a process usually known as
`regraduation'.\footnote{Other notable examples of regraduation include the
Kelvin choice of absolute temperature, the Cox derivation of the sum and
product rule for probabilities, and the derivation of the sum and product
rules for quantum amplitudes. \cite{Caticha 2012} \cite{Caticha 1998}}

A quick examination of eq.(\ref{sch b}) shows that the optimal $k$ is such
that the non-linear term drops out. The optimal choice, which we denote
$\hat{k}$, is
\begin{equation}
\hat{k}=(\frac{\eta^{2}}{8\xi})^{1/2}~.
\end{equation}
We then identify the optimal regraduated $\eta/\hat{k}$ with Planck's constant
$\hbar$,
\begin{equation}
\frac{\eta}{\hat{k}}=(8\xi)^{1/2}=\hbar~,
\end{equation}
and eq.(\ref{sch b}) becomes the linear Schr\"{o}dinger equation,%
\begin{equation}
i\hbar\partial_{t}\Psi=-\frac{\hbar^{2}}{2}m^{AB}\partial_{A}\partial_{B}%
\Psi+V\Psi=%
{\displaystyle\sum\limits_{n}}
\frac{-\hbar^{2}}{2m_{n}}\nabla_{n}^{2}\Psi+V\Psi~, \label{sch c}%
\end{equation}
where the wave function is $\Psi=\rho e^{i\Phi/\hbar}$. The constant
$\xi=\hbar^{2}/8\ $in eq.(\ref{QP a}) turns out to be crucial: it defines the
numerical value of what we call Planck's constant and sets the scale that
separates quantum from classical regimes.

The conclusion is that for any positive value of the constant $\xi$ it is
always possible to regraduate $\Psi_{k}$ to a physically equivalent but more
convenient description where the Schr\"{o}dinger equation is linear. From the
ED perspective the linear superposition principle and the complex Hilbert
spaces are important because they are convenient but not because they are fundamental.

\section{ED in an external electromagnetic field}

In ED the information that is physically relevant for prediction is codified
into constraints that reflect that motion is (a) continuous, (b) correlated
and directional, and (c)\ non-dissipative. These constraints are expressed by
eqs.(\ref{kappa n}), (\ref{kappa prime}) and (\ref{Hamilton b}) respectively.
In this section we show that interactions can be introduced by imposing
additional constraints.

As an explicit illustration we show that the effect of an external
electromagnetic field is modelled by a constraint on the component of
displacements along a certain direction represented by the vector potential
$A_{a}(x)$. For each particle $n$ we impose the constraint
\begin{equation}
\langle\Delta x^{a}\rangle A_{a}(x_{n})=\kappa_{n}^{\prime\prime}%
~,\quad(n=1\ldots N) \label{constraint A}%
\end{equation}
where $\kappa_{n}^{\prime\prime}$ is a particle-dependent constant that
reflects the strength of the coupling to $A_{a}$.

The resemblance between (\ref{constraint A}) and the drift potential
constraint, eq.(\ref{kappa prime}), is very significant --- as we shall see
shortly it leads to gauge symmetries --- but there also are significant
differences. Note that (\ref{kappa prime}) is a single constraint acting in
the $N$-particle configuration space --- it involves the drift potential
$\phi(x)$ with $x\in\mathbf{X}_{N}$. In contrast, (\ref{constraint A}) are $N$
constraints acting in the $1$-particle space --- the vector potential
$A_{a}(x_{n})$ is a function in 3D space, $x_{n}\in\mathbf{X}$.

Except for minor changes the development of ED proceeds as before. The
transition probability $P(x^{\prime}|x)$ that maximizes the entropy
$\mathcal{S}[P,Q]$, eq.(\ref{Sppi}), subject to (\ref{kappa n}),
(\ref{kappa prime}), (\ref{constraint A}) and normalization, is
\begin{equation}
P(x^{\prime}|x)=\frac{1}{\zeta}\exp-%
{\displaystyle\sum\nolimits_{n}}
\left[  \frac{1}{2}\alpha_{n}\delta_{ab}\Delta x_{n}^{a}\Delta x_{n}%
^{b}-\left(  \alpha^{\prime}\frac{\partial\phi}{\partial x_{n}^{a}}-\alpha
_{n}^{\prime\prime}A_{a}(x_{n})\right)  \Delta x_{n}^{a}\right]
\label{trans prob a}%
\end{equation}
which includes an additional set of Lagrange multipliers $\alpha_{n}%
^{\prime\prime}$. Next use eqs.(\ref{alpha n}) and (\ref{mass a}) to get
\begin{equation}
P(x^{\prime}|x)=\frac{1}{\zeta}\exp-\left[  \frac{1}{2\eta\Delta t}%
m_{AB}\Delta x^{A}\Delta x^{B}-\left(  \partial_{A}\phi-\frac{1}{\eta}%
A_{A}\right)  \Delta x^{A}\right]  \label{trans prob b}%
\end{equation}
where we have absorbed $\alpha^{\prime}$ into $\phi$, $\alpha^{\prime}%
\phi\rightarrow\phi$, and written
\begin{equation}
A_{A}(x)=\eta\alpha_{n}^{\prime\prime}A_{a}(x_{n})~
\end{equation}
as a vector in configuration space. As in eq.(\ref{Delta x}) a generic
displacement $\Delta x^{A}$ can be expressed in terms of a expected drift plus
a fluctuation, $\Delta x^{A}=b^{A}\Delta t+\Delta w^{A}$, but the drift
velocity now includes a new term,%
\begin{equation}
b^{A}=m^{AB}\left(  \eta\partial_{A}\phi-A_{A}\right)  ~. \label{drift}%
\end{equation}
The fluctuations $\Delta w^{A}$, eq.(\ref{fluc}), remain unchanged.

A very significant feature of the transition probability $P(x^{\prime}|x)$ is
its invariance under gauge transformations,
\begin{align}
\frac{\partial\phi}{\partial x_{n}^{a}}  &  \rightarrow\frac{\partial
\tilde{\phi}}{\partial x_{n}^{a}}=\frac{\partial\phi}{\partial x_{n}^{a}%
}+\frac{1}{\alpha^{\prime}}\frac{\partial\chi(x_{n})}{\partial x_{n}^{a}}\\
A_{a}(x_{n})  &  \rightarrow\tilde{A}_{a}(x_{n})=A_{a}(x_{n})+\frac{1}%
{\alpha^{\prime\prime}}\frac{\partial\chi(x_{n})}{\partial x_{n}^{a}}~.
\end{align}
Note that these transformations are local in space. (The vector potential
$A_{a}(x_{n})$ and the gauge function $\chi(x_{n})$ are functions in space.)
They can be written in the $N$-particle configuration space,
\begin{align}
\phi &  \rightarrow\tilde{\phi}=\phi+\frac{1}{\eta}\bar{\chi}~,\\
A_{A}  &  \rightarrow\tilde{A}_{A}=A_{A}+\partial_{A}\bar{\chi}~.
\end{align}
where
\begin{equation}
\bar{\chi}(x)=%
{\displaystyle\sum\nolimits_{n}}
\chi(x_{n})~.
\end{equation}

The accumulation of many small steps is described by a Fokker-Planck equation
which can be written either as a continuity equation, eq.(\ref{FP b}), or in
its Hamiltonian form, eq.(\ref{Hamilton a}). As might be expected, the current
velocity $v^{A}$, eq.(\ref{curr}), and the ensemble Hamiltonian $\tilde{H}$,
eq.(\ref{Hamiltonian a}), must be suitably modified,
\begin{equation}
v^{A}=m^{AB}\left(  \partial_{B}\Phi-A_{B}\right)  \quad\text{with}\quad
\Phi=\eta(\phi-\log\rho^{1/2})~, \label{curr A}%
\end{equation}
and%
\begin{equation}
\tilde{H}[\rho,\Phi]=\int dx\,\left[  \frac{1}{2}\rho m^{AB}\left(
\partial_{A}\Phi-A_{A}\right)  \left(  \partial_{B}\Phi-A_{B}\right)  +\rho
V+\,\frac{\hbar^{2}}{8\rho}m^{AB}\partial_{A}\rho\partial_{B}\rho\right]  ~.
\label{Hamiltonian b}%
\end{equation}
As a shortcut here we have adopted the same functional $F[\rho]$ motivated by
information-geometry, eq.(\ref{QP a}), and set $\xi=\hbar^{2}/8$. The new FP
equation now reads,
\begin{equation}
\partial_{t}\rho=-\partial_{A}\left[  \rho m^{AB}\left(  \partial_{B}%
\Phi-A_{B}\right)  \right]  ~. \label{FP d}%
\end{equation}
The requirement that $\tilde{H}$ be conserved for arbitrary initial conditions
amounts to imposing the second Hamilton equation, eq.(\ref{Hamilton b}), which
leads to the Hamilton-Jacobi equation,
\begin{equation}
\partial_{t}\Phi=-\frac{1}{2}m^{AB}\left(  \partial_{A}\Phi-A_{A}\right)
\left(  \partial_{B}\Phi-A_{B}\right)  +V-\frac{\hbar^{2}}{2}m^{AB}%
\frac{\partial_{A}\partial_{B}\rho^{1/2}}{\rho^{1/2}}~~. \label{HJ d}%
\end{equation}
Finally, we combine $\rho$ and $\Phi$ into a single wave function, $\Psi=\rho
e^{i\Phi/\hbar}$, to obtain the Schr\"{o}dinger equation,
\begin{equation}
i\hbar\partial_{t}\Psi_{k}=-\frac{\hbar^{2}}{2}m^{AB}(\partial_{A}-\frac
{i}{\hbar}A_{A})(\partial_{B}-\frac{i}{\hbar}A_{B})\Psi_{k}+V\Psi_{k}~.
\label{sch d}%
\end{equation}

We conclude with two comments. First, the covariant derivative
\begin{equation}
D_{A}=\partial_{A}-\frac{i}{\hbar}A_{A}=\frac{\partial}{\partial x_{n}^{a}%
}-\frac{i}{\hbar}\eta\alpha_{n}^{\prime\prime}A_{a}(x_{n}) \label{D a}%
\end{equation}
in (\ref{sch d}) can be written in the standard notation,
\begin{equation}
D_{A}=\frac{\partial}{\partial x_{n}^{a}}-\frac{ie_{n}}{\hbar c}A_{a}(x_{n})~,
\label{D b}%
\end{equation}
where $e_{n}$ is the electric charge of particle $n$ in units where $c$ is the
speed of light. Comparing (\ref{D a}) with (\ref{D b}) allows us to interpret
the Lagrange multipliers
\begin{equation}
\alpha_{n}^{\prime\prime}=\frac{e_{n}}{\eta c}%
\end{equation}
in terms of the electric charges $e_{n}$ and the speed of light $c$. Thus, in
ED electric charge $e_{n}$ is essentially a Lagrange multiplier $\alpha
_{n}^{\prime\prime}$ that regulates the response to the external
electromagnetic potential $A_{a}$.

The second comment is that the derivation above is limited to static external
potentials, $\partial_{t}V=0$ and $\partial_{t}A_{a}=0$, so that energy is
conserved. This limitation is easily lifted. For time-dependent potentials the
relevant energy condition must take into account the work done by external
sources: we require that the energy increase at the rate
\begin{equation}
\frac{d\tilde{H}}{dt}=\frac{\partial\tilde{H}}{\partial t}~.
\end{equation}
The net result is that equations (\ref{FP d}), (\ref{HJ d}) and (\ref{sch d})
remain valid for time-dependent external potentials.

\section{Some remarks and conclusions}

\paragraph*{Are there any new predictions?}

Our goal has been to derive dynamical laws and in particular quantum theory as
an example of entropic inference. This means that to the extent that we
succeed and derive quantum mechanics and not some other theory we should not
expect predictions that deviate from those of the standard quantum theory ---
at least not in the non-relativistic regime discussed here. However, the
motivation behind the ED program lies in the conviction that it will
eventually allow us to extend it to other realms, such as gravity or
cosmology, where the status of quantum theory is more questionable.

\paragraph*{The Wallstrom objection --}

An important remaining question is whether the Fokker-Planck and the
generalized Hamilton-Jacobi equations, eqs.(\ref{Hamilton a}) and
(\ref{Hamilton b}), are fully equivalent to the Schr\"{o}dinger equation. This
point was first raised by Wallstrom \cite{Wallstrom 1989} in the context of
Nelson's stochastic mechanics \cite{Nelson 1985} and concerns the single- or
multi-valuedness of phases and wave functions. Briefly the objection is that
stochastic mechanics leads to phases $\Phi$ and wave functions $\Psi$ that are
either both multi-valued or both single-valued. Both alternatives are
unsatisfactory: quantum mechanics forbids multi-valued wave functions, while
single-valued phases can exclude physically relevant states (\emph{e.g.},
states with non-zero angular momentum). Here we do not discuss this issue in
any detail except to note that the objection does not apply once particle spin
is incorporated into ED. As shown by Takabayasi \cite{Takabayasi 1983} a
similar result holds for the hydrodynamical formalism. The basic idea is that,
as mentioned earlier, the drift potential $\phi$ should be interpreted as an
angle. Then the integral of the phase $d\Phi$ over a closed path gives
precisely the quantization condition that guarantees that wave functions
remain single-valued even for multi-valued phases.

\paragraph*{Epistemology \emph{vs}. ontology --}

Dynamical laws have been derived as an example of entropic dynamics. In this
model \textquotedblleft reality\textquotedblright\ is reflected in the
positions of the particles, and our \textquotedblleft limited information
about reality\textquotedblright\ is represented in the probabilities as they
are updated to reflect the physically relevant\ constraints.

\paragraph*{Quantum non-locality --}

ED may appear classical because no \textquotedblleft quantum\textquotedblright%
\ probabilities were introduced. But this is not so. Probabilities, in this
approach, are neither classical nor quantum; they are tools for inference.
Phenomena that would normally be considered non-classical, such as non-local
correlations, are the natural result of including the quantum potential term
in the ensemble Hamiltonian.

\paragraph*{On dynamical laws --}

Action principles are not fundamental; they are convenient ways to summarize
the dynamical laws derived from the deeper principles of entropic inference.
The requirement that an energy be conserved is an important piece of
information (\emph{i.e.}, a constraint) which will probably receive its full
justification once a completely relativistic extension of entropic dynamics to
gravity is developed.

\paragraph*{On entropic \emph{vs}. physical time --}

The derivation of laws of physics as examples of inference led us to introduce
the informationally motivated notion of entropic time which includes
assumptions about the concept of instant, of simultaneity, of ordering, and of
duration. It is clear that entropic time is useful but is this the actual,
real, \textquotedblleft physical\textquotedblright\ time? The answer is yes.
By deriving the Schr\"{o}dinger equation (from which we can obtain the
classical limit) we have shown that the $t$ that appears in the laws of
physics is entropic time. Since these are the equations that we routinely use
to design and calibrate our clocks we conclude that \emph{what clocks measure
is entropic time}. No notion of time that is in any way deeper or more
\textquotedblleft physical\textquotedblright\ is needed. Most interestingly,
the entropic model automatically includes an arrow of time.

\paragraph*{Acknowledgments}

I would like to thank D. Bartolomeo, C. Cafaro, N. Caticha, S. DiFranzo, A.
Giffin, P. Goyal, S. Ipek, D.T. Johnson, K. Knuth, S. Nawaz, M. Reginatto, C.
Rodr\'{\i}guez, J. Skilling, and C.-Y. Tseng for many discussions on entropy,
inference and quantum mechanics.

\end{document}